\documentclass[10pt]{article}

\begin{document}

\title{The Effective Energy-Momentum Tensor in Kaluza-Klein Gravity With Large Extra Dimensions and Off-Diagonal Metrics}
\author{J. Ponce de Leon\thanks{E-mail: jponce@upracd.upr.clu.edu or jpdel1@hotmail.com}\\ Laboratory of Theoretical Physics, Department of Physics\\ 
University of Puerto Rico, P.O. Box 23343, San Juan, \\ PR 00931, USA}

\maketitle

\begin{abstract}
We consider a version of Kaluza-Klein theory where the cylinder condition is not imposed. The metric is allowed to have explicit dependence on the ``extra" coordinate(s).  This is the usual scenario in brane-world and space-time-matter theories. We extend the usual discussion by considering five-dimensional metrics with off-diagonal terms. We replace the condition of cylindricity by the requirement that physics in four-dimensional space-time should remain invariant under changes of coordinates in the five-dimensional bulk. This invariance does not eliminate physical effects from the extra dimension but separates them from spurious geometrical ones. We use the appropriate splitting technique to construct the most general induced energy-momentum tensor, compatible with the required invariance. It generalizes all previous results in the literature. In addition, we find two four-vectors, $J_{m}^{\mu}$ and $J_{e}^{\mu}$, induced by off-diagonal metrics, that separately satisfy the usual equation of continuity in $4D$. These vectors appear as source-terms in equations that closely resemble the ones of electromagnetism. These are Maxwell-like equations for an antisymmetric tensor ${\hat F}_{\mu\nu}$ that generalizes the usual electromagnetic one. This generalization is not an assumption, but follows naturally from the dimensional reduction. Thus, $if$ ${\hat F}_{\mu\nu}$ could be identified with the electromagnetic tensor, then the theory $would$ predict  the existence of classical magnetic charge and current. The splitting formalism used allows us to construct $4D$ physical quantities from five-dimensional ones, in a way that is independent on how we choose our space-time coordinates from those of the bulk. 

 \end{abstract}

 
\newpage

\section{Introduction}
Most of the recent advances in theoretical physics deal with models of our universe in more than four dimensions. Theories of the Kaluza-Klein type in many dimensions are used in different branches of physics. Superstrings ($10D$) and supergravity $(11D)$ are well known examples.

 In gravitation and cosmology the study of multidimensional theories called ``brane world" models as well as the space-time-matter (STM) theory in $5D$, which seeks the unification of matter and geometry, have become quite popular. Although, these  theories have different physical motivation for the introduction of a large extra dimension, they share the same goals and philosophy \cite{STM-Brane}.

The crucial question in these theories can be split up in two parts. (1) How to extract the correct $4D$ interpretation from geometrical objects, like scalar dilatonic fields and antisymmetric forms, which appear in more than four dimensions? (2) How to predict observable effects from the extra dimensions? It seems to be obvious that the successful answer to the second question requires the correct answer to the first one. 
\subsection{Large extra dimension}
 This paper deals with the first part of this question. We consider a general version of Kaluza-Klein theory in $5D$ where the extra dimension is $not$ assumed to be compactified. From a mathematical viewpoint, this means that the $5$-dimensional metric tensor is allowed to depend explicitly on the fifth coordinate. This is the common assumption made in brane-world theory and STM \cite{STM-Brane}-\cite{Deruelle and Katz}.

 However, we extend the discussion to the case where the five-dimensional metrics possess non-vanishing off-diagonal terms. The concrete question that we ask here is the following: What is the more general expression for an energy-momentum tensor in four-dimensions, that can be constructed with the elements of the $5D$ theory (The discussion can be easily extended to any number of dimensions. In this case the question is about the energy-momentum tensor constructed in one lower dimension).

The appearance of the extra dimensions in $4D$ physical laws is usually avoided by imposing the ``cylindricity condition", which essentially requires that all derivatives with respect to the extra coordinate(s) vanish. However, this condition is too restrictive. Indeed, this condition can be replaced by a less stringent one. Namely the requirement that physics in $4D$ should be invariant under coordinate transformations in $5D$ (to which we will refer as gauge invariance). Although cylindricity always implies gauge invariance, this invariance does not necessarily require cylindricity.

Therefore, in this work we require gauge invariance of $4D$ physical quantities instead of cylindricity. In a recent work \cite{recent work} we applied the same requirement to study the motion of test particles in Kaluza-Klein Gravity. Gauge invariance allowed us to clarify a number of aspects of the so called ``fifth force". In particular, we provided a new definition for this force where it is gauge invariant and orthogonal to the four-velocity of the particle \cite{Phys.Lett}.   
\subsection{Previous results}
Restricted versions of the problem under consideration here are well known in the literature. 

The first version is the ``classical"  Kaluza-Klein theory, where the metric in $5D$ is restricted by the cylinder condition. This means that the extra dimension is assumed to be compactified. In this case, the five-dimensional Einstein equations, in vacuum, may be reduced to ones that resemble the $4D$ equations of general relativity in the presence of an electromagnetic field \cite{Nodvik}-\cite{Appelquist and Chodos}. Here, the dimensional reduction is quite simple; in practice it amounts to integrating over the extra variable in the action integral. 

The second version \cite{Wesson and JPdeL} relaxes cylindricity, i.e., the extra dimension is $not$ assumed to be compactified. As a consequence, a simple integration in the action is no longer possible and the dimensional reduction is much more involved than in the classical case.  However, a great simplification is attained if one restricts the discussion to the case where $\gamma_{\mu 4}=0$\footnote{In this paper $\gamma_{AB}$ denotes the five-dimensional metric}. This is the usual working scenario in brane-world and STM theories \cite{STM-Brane}. This restriction amounts to describing a space-time free of electromagnetic field.  The induced $4D$ energy-momentum tensor presents reasonable physical properties. In the context of brane-world theory, it predicts local and non-local five-dimensional corrections to general relativity. In STM, it constitutes the basis for the geometrical interpretation of matter \cite{Overduin}. As an important illustration of this, we mention that the $5D$ equations, under the appropriate symmetry conditions, lead to Friedmann-Robertson-Walker cosmological models with perfect fluid source, and satisfying the barotropic equation of state $\rho = np$ \cite{JPdeL 1}. 

Both restricted versions use the tacit assumption (commonly seen in the literature) that the first four coordinates correspond to the physical space-time, while the remaining one is the extra  dimension. The obvious question is: given an arbitrary $5$-dimensional metric, how do we know, {\it a priori}, which one is the ``extra" coordinate? What if we choose our space-time coordinates in some another way, say some combination (or function) of the bulk coordinates for example? 

\subsection{General case: off-diagonal metrics} 

Our work is free of these assumptions. In this paper we construct the most general ``induced" energy-momentum tensor compatible with gauge invariance, instead of cylindricity, and $\gamma_{\mu 4} \neq 0$. 

Certainly,  making a change of coordinates in $5D$, one can always transform an  arbitrary metric $\gamma_{AB}$ with $\gamma_{\mu 4} \neq 0$ into another metric, say ${\bar{\gamma}}_{AB}$, with vanishing off-diagonal terms. From a mathematical point of view the ``old" metric and the transformed one    represent the same five-dimensional manifold.

However, they are {\em not} equivalent from a ``physical" point of view. Indeed, they give rise to different scenarios in $4D$. In the first case ($\gamma_{AB}$, $\gamma_{\mu 4} \neq 0$) spin one fields are present since $\gamma_{\mu 4}$ is just such a field. In the second case, the transformed metric ${\bar{\gamma}}_{AB}$, ${\bar{\gamma}}_{\mu 4} = 0$ removes it from the outset. In general, the structure and the material content of the observed spacetime is changed by any coordinate transformation that involves the extra coordinate. There  is a similar situation in four-dimensional GR, where a coordinate transformation involving the time coordinate also implies a  change in the system of reference, and consequently modifies  the observed  physical picture.

Thus, the consideration of $\gamma_{\mu 4} \neq 0$ corresponds to a physical scenario distinct from those  mentioned in section $1.2$. Namely, it allows the presence of fields of spin one in a full theory of Kaluza-Klein with zero and massive modes of graviton. Besides, the examination of the most general metric in $5D$ is important because we do not know yet which metric frame is the correct representation of our four-dimensional spacetime.

Therefore, we provide here a method to obtain physical quantities from $5D$ that is independent of how we choose our spacetime coordinates. The dimensional reduction for this general version generalizes all previous results in the literature. In addition, we find two conserved four-vectors, which appear as source-terms in equations that closely resemble those of electromagnetism. In this analogy, one of them corresponds to an effective electrical charge and current, while the other $4$-vector would correspond an effective $magnetic$  charge and current. 

The paper is organized as follows. In Section 2, we consider the $5D$ line element and field equations and provide some details of our splitting technique in an arbitrary set of coordinates. In Section 3, we do the splitting of the field equations. They are expressed in terms of three quantities only. These are well defined, gauge invariant, $4D$ quantities constructed from the elements of the $5D$ theory. In Section 4, we introduce the {\em physical metric} and construct  the energy-momentum tensor as measured by an observer in physical $4D$ space-time. In Sections 5 and 6, we introduce what we call the {\em coordinate frame}  and show that our general approach contains all previous results in the literature. In Section 7, we discuss the fully general case. We will see that, as a consequence, of the anholonomic nature of the theory, there is a quantity that {\it can} be interpreted as magnetic charge and current. It is represented by a four-vector that satisfies the usual continuity equation. Finally, in Section 8, we present a summary of our conclusions.    

\section{General Framework}
In this section we will show how to construct four-dimensional quantities from five-dimensional ones. The construction is independent on the specific coordinates used in the $5D$ manifold and the way we choose coordinates in $4D$. Then we introduce the {\it local frame}, which is the frame attached to an observer in $4D$.

We will consider a general five-dimensional manifold with coordinates $\xi^{A}$ $(A = 0,1,2,3,4)$ and metric tensor $\gamma_{AB}(\xi^{C})$. The $5D$ interval is then given by
\begin{equation}
\label{general 5D metric}
d{\cal S}^2 = \gamma_{AB}d{\xi}^A d{\xi}^B.
\end{equation}

The field equations are taken to be the $5D$ Einstein equations,
\begin{equation}
\label{Field equations in 5D} 
{\cal R}_{AB} - \frac{1}{2}\gamma_{AB}{\cal R} = k {\cal T}_{AB},\;\; (k = const.)
\end{equation}
where ${\cal R}_{AB}$ is the Ricci tensor constructed in terms of the Christoffel symbols  

\begin{equation}
\label{Christoffel Symbols in 5D}
{\cal K}^{A}_{BC} = \frac{1}{2}\gamma^{AD}\left(\gamma_{DB,C} + \gamma_{DC,B} - \gamma_{BC,D}\right),
\end{equation}
${\cal R} = \gamma^{AB}{\cal R}_{AB}$ is the five-dimensional scalar curvature, and ${\cal T}_{AB}$ a five-dimensional energy-momentum tensor.

We have already mentioned that are two cases where the $(4 + 1)$ separation of these equations is quite straightforward. In the first case the metric is general, except for the restriction that it does not depend explicitly on the extra coordinate. In the second case the metric is allowed to depend on the extra coordinate but it is restricted by the assumption $\gamma_{\mu4} = 0$. In both cases, the assumption that the first four coordinates represent the physical space-time, allows to obtain the $4D$ part of (\ref{Field equations in 5D}) and (\ref{Christoffel Symbols in 5D}) by simply putting $A, B = 0,1,2,3$. 

However, this simple method {\em does not} work in the general case under consideration here. Indeed, the $4D$ quantities defined with this straightforward identification (that $A, B = 0,1,2,3$ are the space-time components of (\ref{Field equations in 5D}) and (\ref{Christoffel Symbols in 5D})) {\em are not invariant under transformations in} $5D$ \cite{recent work}-\cite{Phys.Lett}.
 
Thus, in the general case where $\gamma_{4\mu} \neq 0$ and the metric is allowed to depend on the extra variable, another splitting method should be used in order to assure gauge invariance of physical quantities.

 In this paper we use the ``local" frame of reference, introduced in a previous work \cite{recent work}, to reduce the five-dimensional equations and construct the physical quantities in $4D$ that are invariant under  transformations in $5D$. 
\subsection{Splitting technique}
To facilitate the discussion and set the notation, we start with a brief summary of our splitting technique. For details we refer the reader to Ref. \cite{recent work} We assume that the $5D$ manifold in (\ref{general 5D metric}) allows us to construct, appropriately (see bellow), a four-dimensional hypersurface that can be identified with our $4D$ space-time. In this hypersurface we introduce a set of four parameters $x^{\mu}$ $(\mu= 0,1,2,3)$, which are functions of $\xi^A$,
\begin{equation}
\label{4D coordinates}
x^{\mu}= x^{\mu}\left({\xi^0},{\xi^1},{\xi^2},{\xi^3},{\xi^4}\right).
\end{equation}
The derivatives of these functions with respect to $\xi^A$
\begin{equation}
\label{Basis in 4D}
\hat{e}^{(\mu)}_{A}= \frac{\partial{x^{\mu}}}{\partial{\xi^A}},
\end{equation}
behave as covariant vectors\footnote{The index in parenthesis numbers the vector, while the other one indicates its coordinate in $5D$.} with respect to changes $\xi^A = {\xi^A}(\bar{\xi}^B)$ in $5D$, and as  contravariant vectors with respect to transformations $x^{\mu}= x^{\mu}(\bar{x}^{\nu})$ in $4D$. At each point these vectors are tangent to the hypersurface. Therefore, in the region where they are linearly independent, they constitute a basis for the $4D$ hypersurface under consideration. We will interpret this, appropriately defined $4D$ manifold, as the physical space-time and $x^{\mu}$ as the coordinates in it.

We can now introduce the vector $\psi^{A}$, orthogonal to space-time. This is completely determined by 
\begin{eqnarray}
\label{definition of psi}
\hat{e}^{(\mu)}_{A}\psi^{A}&=&0,\nonumber \\
\gamma_{AB}\psi^A\psi^B &=& \epsilon,
\end{eqnarray}
where $\epsilon$ is taken to be $+1$ or $-1$ depending on whether the extra dimension is timelike or spacelike, respectively. We will also need the set of vectors $\hat{e}_{(\mu)}^{A}$, defined  as
\begin{eqnarray}
\label{associate basis}
\hat{e}_{A}^{(\mu)}\hat{e}^{A}_{(\nu)}&=& \delta^{\mu}_{\nu},\nonumber \\     
\psi_{A}\hat{e}^{A}_{(\mu)}&=& 0.
\end{eqnarray} 
They allow us to express any infinitesimal displacement in $5D$ as
\begin{equation}
\label{5D displacement}
d\xi^A = \hat{e}^{A}_{(\mu)}dx^{(\mu)}+ \epsilon \psi^A dx^{(4)},
\end{equation}
 where $dx^{(\mu)}= \hat{e}_{B}^{(\mu)}d\xi^B$ and $dx^{(4)} = \psi_{B}d\xi^B$ represent the displacements along the corresponding basis vectors. From this, it follows that 
\begin{equation}
\label{projector}
\hat{e}_{(\mu)}^{A}\hat{e}^{(\mu)}_{B}= \delta_{B}^{A}- \epsilon \psi^{A}\psi_{B},
\end{equation}

\subsection{Local frame}
Thus, in the neighborhood of each point the observer is armed with five independent vectors, $\hat{e}^{(\mu)}_{A}$ and $\psi^{B}$, that constitute its frame of reference. In order to simplify further calculations we introduce a more symmetrical notation. To this end we set 
\begin{eqnarray}
\label{extra basis vector}
\hat{e}^{(4)}_{A}&=& \psi_{A},\nonumber  \\
\hat{e}^{A}_{(4)}&=& \epsilon \psi^A.
\end{eqnarray}
Now, Eqs. (\ref{definition of psi}), (\ref{associate basis}) and (\ref{projector}) become
\begin{eqnarray}
\label{general relation between basis vectors}
\hat{e}_{A}^{(B)}\hat{e}^{A}_{(C)}&=& \delta_{C}^{B},\nonumber  \\
\hat{e}^{(N)}_{C}\hat{e}_{(N)}^{D}&=& \delta_{C}^{D}.
\end{eqnarray}
Finally, we define our ``local" $5D$ metric $\hat{g}_{(A)(B)}$ as
\begin{equation}
\label{Local 5D metric}
\hat{g}_{(A)(B)}= \gamma_{MN}\hat{e}_{(A)}^{M}\hat{e}_{(B)}^{N}.
\end{equation}
which neatly separates the five-dimensional manifold in $(4 + 1)$ parts, viz.,
\begin{equation}
\label{local frame}
\hat{g}_{(A)(B)}= \bordermatrix{  &  &      \cr
                                  &\hat{g}_{\mu\nu} &0    \cr
                                  &0 &\epsilon    \cr},
\end{equation}
where $\hat{g}_{\mu\nu}$ is the metric (to which we will refer as the {\it induced} $4D$  metric\footnote{Later, in Section 4.1 we will define the {\it physical} or {\it observable} $4D$ {\it metric} $g_{\mu\nu}$ as being conformally related to the induced one.}) of the four-dimensional space-time section, viz.,
\begin{equation}
\label{4D metric}
\hat{g}_{\mu\nu}= \hat{e}_{(\mu)}^{A}\hat{e}_{(\nu)}^{B}\gamma_{AB}.
\end{equation}

The advantage of the local frame is that it provides a $(4+1)$ separation which is invariant under arbitrary changes of coordinates in $5D$. Namely, putting $(A), (B) = 0, 1,2,3$ we obtain the space-time part of the local metric. We will use this frame, in the next Section, to perform the separation of the five-dimensional equations (\ref{Field equations in 5D}).

The $5D$ interval becomes
\begin{equation}
\label{5D interval in local metric}
d{\cal S}^2 = \hat{g}_{(A)(B)}dx^{(A)}dx^{(B)}.
\end{equation}
In addition, from (\ref{general relation between basis vectors}), it follows that 
\begin{equation}
\label{coordinate metric from local metric}
\gamma_{AB}= \hat{e}_{A}^{(C)}\hat{e}_{B}^{(D)}\hat{g}_{(C)(D)}.
\end{equation}
We also notice
\begin{equation}
\gamma^{AB} = \hat{e}^{A}_{(P)}\hat{e}^{B}_{(Q)}\hat{g}^{(P)(Q)},
\end{equation}
where 
\begin{equation}
\hat{g}^{(P)(Q)}= \bordermatrix{  &  &      \cr
                                  &\hat{g}^{\alpha\beta} &0    \cr
                                  &0 &\epsilon    \cr},
\end{equation}
 
Finally, we mention that basis indexes are lowered and raised with $ \hat{g}_{(A)(B)}$, while  $5D$ coordinate indexes are lowered and raised with $\gamma_{AB}$.

\section{Field Equations In The Local Frame}
In this Section we obtain the explicit form of the field equations in an arbitrary local frame. They neatly separate in three sets of 1, 4 and 10 equations. These equations are expressed solely in terms of well defined, gauge invariant, $4D$ quantities constructed from the $5D$ theory.

In the local frame, the $5D$ line element becomes 
\begin{equation}
d{\cal S}^2 = \hat{g}_{(\mu)(\nu)}dx^{(\mu)}dx^{(\nu)} + \epsilon (dx^{(4)})^2.
\end{equation}
The field equations (\ref{Field equations in 5D}), using (\ref{coordinate metric from local metric}), can be written as
\begin{equation}
\label{field equations in the local frame}
\hat{\cal R}_{(A)(B)} = k\left(\hat{\cal T}_{(A)(B)} - \frac{1}{3}\hat{g}_{(A)(B)}\hat{\cal T}\right),
\end{equation}
where 
\begin{equation}
\hat{\cal R}_{(A)(B)}= \hat{e}_{(A)}^{P}\hat{e}_{(B)}^{Q}{\cal R}_{PQ},
\end{equation}
are the components of the five-dimensional Ricci tensor on the local frame. $\hat{\cal T}_{(A)(B)}= \hat{e}_{(A)}^{P}\hat{e}_{(B)}^{Q}{\cal T}_{PQ}$ and $\hat{\cal T}$ are, respectively, the basis components of the $5D$ energy-momentum tensor and its corresponding trace.

Now, since the local frame provides a clear separation between the space-time and the extra part of the metric, setting $A\rightarrow \mu$ and $B\rightarrow \nu$ in $\hat{\cal R}_{(A)(B)}$ we obtain the $4D$ part of this $5D$ tensor. 

It may be worth emphasizing that there is no transformation of coordinates, neither in $5D$ nor in $4D$, relating the above line element in the local frame to the general $5D$ metric (\ref {general 5D metric}). So although the line element is diagonal in the local frame, we have not eliminated the spin-one field. The simplicity of the above method  for extracting the $4D$ quantities is one of the major advantages of using the local frame. 
 
\subsection{Christoffel symbols}
The first step, in order to find the basis components $\hat{\cal R}_{(A)(B)}$, is to obtain an appropriate expression for the ``coordinate" Christoffel symbols (\ref{Christoffel Symbols in 5D}) in terms of basis vectors. For this we substitute (\ref{coordinate metric from local metric}) into (\ref{Christoffel Symbols in 5D}). After some manipulations, we get
\begin{equation}
\label{projected Christoffel symbols}
{\cal K}^{D}_{AC} = \hat{e}_{A}^{(P)}\hat{e}_{C}^{(Q)}\hat{e}^{D}_{(R)}\left(\hat{K}^{(R)}_{(P)(Q)} + \hat{V}^{(R)}_{(P)(Q)}\right),
\end{equation}
where $\hat{K}^{(R)}_{(P)(Q)}$ are the Christoffel symbols constructed with the local metric, viz.,
\begin{equation}
\label{Christoffel symbols constructed with the local metric}
\hat{K}^{(R)}_{(P)(Q)} = \frac{1}{2}\hat{g}^{(R)(S)}\left(\hat{g}_{(S)(P)|(Q)} + \hat{g}_{(S)(Q)|(P)} - \hat{g}_{(P)(Q)|(S)}\right),
\end{equation}
and
\begin{equation}
\hat{V}^{(R)}_{(P)(L)} = \frac{1}{2}\hat{g}^{(R)(Q)}\left(\hat{g}_{(4)(L)}{\cal F}_{(P)(Q)}^{(4)} + \hat{g}_{(4)(P)}{\cal F}_{(L)(Q)}^{(4)}\right) - \frac{1}{2}S^{(R)}_{(P)(L)}.
\end{equation}
Here we have introduced the antisymmetric and symmetric quantities ${\cal F}^{(M)}_{(P)(Q)}$ and $S^{(M)}_{(P)(Q)}$, as
\begin{equation}
\label{antisymmetric tensor}
{\cal F}^{(M)}_{(P)(Q)} = \hat{e}^{(M)}_{N}(\hat{e}^{N}_{(P)|(Q)} - \hat{e}^{N}_{(Q)|(P)}),
\end{equation}
\begin{equation}
\label{symmetric tensor}
S^{(M)}_{(P)(Q)} = \hat{e}^{(M)}_{N}(\hat{e}^{N}_{(P)|(Q)} + \hat{e}^{N}_{(Q)|(P)})
\end{equation}
In Eqs. (\ref{Christoffel symbols constructed with the local metric})-(\ref{symmetric tensor}) and in what follows, the symbol $``|"$ indicates directional derivatives. Specifically, for an arbitrary quantity $\phi$ we have
\begin{equation}
\label{definition of projected derivatives}
\phi_{|(Q)} = \frac{\partial \phi}{\partial x^{(Q)}} = \hat{e}^{A}_{(Q)}\frac{\partial \phi}{\partial \xi^{A}}.
\end{equation}
Reciprocally, 
\begin{equation}
\label{reciprocal relation}
\frac{\partial \phi}{\partial \xi^A} = \phi_{|(M)}{\hat{e}}^{(M)}_{A},
\end{equation}

The antisymmetric nature of ${\cal F}^{(M)}_{(P)(Q)}$ implies that ${\cal F}^{(\lambda)}_{(P)(Q)} = 0$. Indeed, setting $M \rightarrow \lambda$ in (\ref{antisymmetric tensor}) and using orthogonality conditions (\ref{general relation between basis vectors}) in (\ref{antisymmetric tensor}) we obtain
\begin{equation}
\label{space part of antisymmetric tensor}
{\cal F}^{(\mu)}_{(A)(B)} = \hat{e}^{P}_{(B)} \hat{e}^{(\mu)}_{P|(A)}- \hat{e}^{P}_{(A)}\hat{e}^{(\mu)}_{P|(B)}.
\end{equation}
We now need to remember that $\hat{e}^{(\mu)}_{A}= (\partial{x^\mu}/{\partial{\xi^A}})$. Using this, and since $({\partial}^2/{\partial{\xi}^P}{\partial{\xi^Q}})= ({\partial}^2/{\partial{\xi}^Q}{\partial{\xi^P}})$, it follows that
\[
{\cal F}^{(\lambda)}_{(A)(B)} = 0. \]

 Consequently, only the ten remaining components ${\cal F}^{(4)}_{(P)(Q)}$ are different from zero, in general. In a previous work (equations $(78)$ and $(79)$ in Ref.\cite{recent work}), we have seen that six of them are associated with the electromagnetic field, while the other four with the so called ``fifth" force. In what follows the index $``(4)"$ in ${\cal F}^{(4)}_{(P)(Q)}$ will be suppressed. 

Let us immediately notice that ${\cal{F}}_{(A)(B)}$ and $S^{(C)}_{(A)(B)}$ have distinct behavior under coordinate transformations $\xi^A = \xi^A({\bar{\xi}}^B)$ in $5D$. Indeed, under such transformations ${\hat{e}}^{N}_{(P)}$ is a five-vector, but ${\hat{e}}^{N}_{(P)|(Q)}$ is not. Namely, 
\begin{equation}
\label{trans prop of F and S}
\bar{{\hat{e}}}^{N}_{(P)|(Q)} = \frac{\partial^2 {\bar{\xi}}^N}{\partial \xi^A \partial \xi^B} {\hat{e}}^{A}_{(P)}{\hat{e}}^{B}_{(Q)} + \frac{\partial \bar{\xi}^N}{\partial \xi^ A}e^{A}_{(P)|(Q)}.
\end{equation}
Therefore, from (\ref{antisymmetric tensor}) we find that ${\cal F}^{(M)}_{(P)(Q)}$ remains {\em invariant} under the general transformations  $\xi^A = \xi^A({\bar{\xi}}^B)$. On the other hand, from (\ref{symmetric tensor}) we see that in new coordinates the quantity $S^{(C)}_{(A)(B)}$  has an additional term proportional to $(\partial^2 {\bar{\xi}}^N /\partial \xi^A \partial \xi^B)$, which vanishes only for lineal transformations. Therefore, $S^{(C)}_{(A)(B)}$ is {\em not} invariant under general transformations in $5D$.
  
For completeness, we now proceed to show how the quantities $S^{(C)}_{(A)(B)}$ and ${\cal{F}}_{(A)(B)}$ are related to the second derivatives $(\partial^2\phi/\partial \xi^A \partial \xi^B)$ and $\phi_{|(M)(N)}$, for an arbitrary function $\phi$. The derivative 
\begin{equation}
\frac{\partial}{\partial \xi^B}\left(\frac{\partial \phi}{\partial \xi^A}\right) = \left(\phi_{|(M)}{\hat{e}}^{(M)}_{A}\right)_{|(N)}{\hat{e}}^{(N)}_{B},
\end{equation}
can be rearranged as follows,
\begin{equation}
\frac{\partial^2 \phi}{\partial \xi^B \partial \xi^A} {\hat{e}}^{A}_{(M)}{\hat{e}}^{B}_{(N)} = \phi_{|(N)(M)} - \phi_{|(P)}{\hat{e}}^{(P)}_{C}{\hat{e}}^{C}_{|(N)(M)}.
\end{equation}
where we have used the orthogonality conditions. Now, changing $A \leftrightarrow B$ and $M \leftrightarrow N$ we get
\begin{equation}
\frac{\partial^2 \phi}{\partial \xi^A \partial \xi^B} {\hat{e}}^{A}_{(M)}{\hat{e}}^{B}_{(N)} = \phi_{|(M)(N)} - \phi_{|(P)}{\hat{e}}^{(P)}_{C}{\hat{e}}^{C}_{|(M)(N)}
\end{equation}
If we add these two equations, we obtain
\begin{equation}
\label{symmetry relation}
\frac{\partial^2 \phi}{\partial \xi^A \partial \xi^B} = \frac{1}{2} \left[\left(\phi_{|(M)(N)} + \phi_{|(N)(M)}\right) - \phi_{|(P)}S^{(P)}_{(M)(N)}\right]e^{(M)}_{A}e^{(N)}_{B}.
\end{equation}
On the other hand, if we subtract we obtain 
\begin{equation}
\label{conmutativity relations}
\phi_{|(N)(M)} - \phi_{|(M)(N)} = \phi_{|(P)}{\cal{F}}^{(P)}_{(N)(M)} = \phi_{(4)}{\cal{F}}_{(N)(M)}.
\end{equation}
These expressions are useful in intermediate calculations.

 \subsection{Ricci tensor}
The next step consists of substituting (\ref{projected Christoffel symbols}) into the usual definition for the Ricci tensor ${\cal R}_{AB}$. The calculations involve second derivatives of the metric, which in the local frame become directional derivatives. These derivatives do not commute, in general\footnote{It may be worth mentioning that, although with different purposes, anholonomic frames have recently been considered for the study of off-diagonal metrics in Einstein and brane gravity \cite{anh1}-\cite{Vacaru}}, but the use of relations (\ref{symmetry relation}) and (\ref{conmutativity relations}) keep the  symmetry of the Ricci tensor. 

The calculations are quite long and involved. However, the final result is rather simple. It is expressed in terms of three quantities only. Namely; 
(i) the four-dimensional Ricci tensor $\hat{R}_{\mu\nu}$ constructed with the Christoffel symbols obtained from the induced $4D$ metric, viz.,
\begin{equation}
\label{4D Christoffel symbols}
\hat{\Gamma}^{\lambda}_{\mu\nu} = \frac{1}{2}\hat{g}^{\lambda\sigma}\left(\hat{g}_{\sigma\mu|(\nu)} + \hat{g}_{\sigma\nu|(\mu)} - \hat{g}_{\mu\nu|(\sigma)}\right) = \hat{K}^{(\lambda)}_{(\mu)(\nu)}.
\end{equation}
(ii) the derivatives of the induced $4D$ metric along the fifth dimension, which we denote as
\begin{equation}
\label{definition of Psi}
\hat{\Psi}_{\mu\nu} = \frac{1}{2}\hat{g}_{\mu\nu|(4)},
\end{equation}
and (iii) the antisymmetric tensor ${\cal F}_{(A)(B)}^{(C)}$ defined in (\ref{antisymmetric tensor}). 

Thus, omitting intermediate calculations, we have
\begin{eqnarray}
\label{Ricci tensor in the local basis}
\hat{\cal R}_{(4)(4)}&=& - (\hat{\Psi}_{|(4)} + \hat{\Psi}_{\lambda\rho}\hat{\Psi}^{\lambda\rho}) + \frac{1}{4}{\cal F}_{(\lambda)(\rho)}{\cal F}^{(\lambda)(\rho)} - \epsilon (\hat{D}_{(\lambda)}{\cal F}^{(\lambda)} + {\cal F}_{(\lambda)}{\cal F}^{(\lambda)}),\nonumber \\
\hat{\cal R}_{(\mu)(4)} &=& \hat{D}_{(\lambda)}(\hat{\Psi}_{\mu}^{\lambda} - \delta_{\mu}^{\lambda}\hat{\Psi} + \frac{\epsilon}{2}{\cal F}_{(\mu)}\;^{(\lambda)}) + \epsilon {\cal F}_{(\mu)}\;^{(\rho)}{\cal F}_{(\rho)},\nonumber \\
\hat{\cal R}_{(\mu)(\nu)} &=& \hat{R}_{\mu\nu} - \epsilon(\hat{\Psi}_{\mu\nu|(4)} - 2 \hat{\Psi}_{\mu\rho}\hat{\Psi}^{\rho}_{\nu} + \hat{\Psi}\hat{\Psi}_{\mu\nu}) - \frac{1}{2}(\hat{D}_{(\mu)}{\cal F}_{(\nu)} + \hat{D}_{(\nu)}{\cal F}_{(\mu)})\nonumber \\
 &-& {\cal F}_{(\mu)}{\cal F}_{(\nu)} - \frac{\epsilon}{2}{\cal F}_{(\mu)(\rho)}{\cal F}_{(\nu)}\;^{(\rho)}.
\end{eqnarray}
Here ${\cal F}_{(\mu)(\nu)} = {\cal F}_{(\mu)(\nu)}^{(4)}$; ${\cal F}_{(\mu)} = {\cal F}_{(\mu)(4)} = {\cal F}^{(4)}_{(\mu)(4)}$ . $\hat{D}$ is the operator of covariant differentiation calculated with the $4D$ Christoffel symbols\footnote{The parenthesis in the index of the covariant derivative is kept in order to remind us that it involves the use of directional derivatives (\ref{definition of projected derivatives}). Thus, $\hat{D}_{(\rho)}V_{\lambda} = V_{\lambda|(\rho)} - \hat{\Gamma}^{\sigma}_{\lambda\rho}V_{\sigma}$.} (\ref{4D Christoffel symbols}). Thus, $\hat{D}_{(\lambda)}\hat{g}_{\mu\nu}= 0$. In (\ref{Ricci tensor in the local basis}) indexes are lowered and raised by the induced metric $\hat{g}_{\mu\nu}$. $\hat{\Psi} = \hat{g}^{\mu\nu}\hat{\Psi}_{\mu\nu}$, and $\hat{\Psi}^{\mu\nu} = - (1/2)\left(\hat{g}^{\mu\nu}\right)_{|(4)}$, which follows from (\ref{definition of Psi}) and $\hat{g}^{\mu\nu}\hat{g}_{\nu\lambda} =\delta^{\mu}_{\lambda}$.

The above equations (\ref{Ricci tensor in the local basis}) are totally general. For their derivation we have made no assumptions; neither on the character of the metric nor imposed restrictions on the frame of basis vectors used. In addition, they are (by construction) invariant under general transformations of coordinates in $5D$. We note that the symmetric quantity $S_{(A)(B)}^{(C)}$, which is {\em not} invariant appears nowhere in the final result, as one expected. 

\section{Interpretation In Four-Dimensions}

As we have already mentioned in the Introduction, one of the crucial questions in discussing possible observational manifestations of the extra dimension is: how to give the correct four-dimensional interpretation to the geometrical quantities that appear naturally in theories formulated in more than four dimensions? 

In our five-dimensional case these quantities are: the antisymmetric tensor ${\cal F}_{(\mu)(\nu)}$, the four-vector ${\cal F}_{(\mu)}$ and the symmetric tensor $\hat{\Psi}_{\mu\nu}$. The first two  generalize, to models with a {\em large} extra dimension,  the essential elements of classical Kaluza-Klein theory. Namely the electromagnetic tensor and the gradient of the scalar field, respectively. On the other hand, the quantity $\hat{\Psi}_{\mu\nu}$ is the extrinsic curvature of the four-dimensional subspace representing our universe in brane-world and STM. Therefore it is related to ordinary matter.  

 The above suggests that ${\cal F}_{(\mu)(\nu)}$, ${\cal F}_{(\mu)}$ and  $\hat{\Psi}_{\mu\nu}$  are the, physically relevant, appropriate quantities to construct an   effective four-dimensional energy-momentum which is gauge invariant and consistent with what we already know.

\subsection{Four-dimensional physical metric} 

Firstly, we have to decide about another fundamental question, namely: how to identify the physical or observable space-time metric (which we will denote as $g_{\mu\nu}$) from the induced one?. 

The simplest choice is to assume $\hat{g}_{\mu\nu} = g_{\mu\nu}$. However, this choice may not describe all possible physical scenarios\footnote{In particular, in the context of STM, the identification $\hat{g}_{\mu\nu} = g_{\mu\nu}$, together with the restriction of independence of the fifth coordinate,  lead to a radiation-like equation of state for the induced matter. Also, for static spherically symmetric solutions they lead to an effective energy-momentum tensor which is anisotropic in a number of cases \cite{Wesson book}, \cite{JPdeL 2}, \cite{Wesson and Ponce de Leon 1994}}. In addition, there remains the freedom of multiplying this metric by a conformal factor, namely, 
\begin{equation}
\label{physical metric}
\hat{g}_{\mu\nu} = e^{2\beta}g_{\mu\nu}
\end{equation}
where $\beta$ depends on the dilatonic fields present in the theory. In compactified Kaluza-Klein theory the necessity and uniqueness of the conformal factor has been widely discussed in many papers \cite{Appelquist}-\cite{Bronnikov and Melnikov}.

In this work we will adopt (\ref{physical metric}), which for $\beta = 0$ reduces to the case $\hat{g}_{\mu\nu} = g_{\mu\nu}$. Our main motivation for this choice is to add a degree of freedom  which  we can utilize, in specific applications of the theory, to guarantee that the induced energy-momentum tensor satisfies appropriate physical conditions. Moreover, in order to keep the generality of the discussion, and for future references, we allow $\beta$ to be an arbitrary {\em scalar} function of the space-time coordinates and the extra variable. This assumption, clearly, does not affect the $4 + 1$ separation. 
 
Physical predictions, in particular the interpretation of the matter content of the space-time as well as the behavior of test particles, crucially depend on the choice of $\beta$. 

Mathematically the transition from the induced metric to the physical one is quite simple. It amounts the replacement in the field equations of ``induced" quantities (i.e., calculated with $\hat{g}_{\mu\nu}$) by ``physical" ones (calculated with $g_{\mu\nu}$). Christoffel symbols are obtained by substituting (\ref{physical metric}) into (\ref{4D Christoffel symbols})
\begin{equation}
\hat{\Gamma}^{\mu}_{\lambda\rho} = \Gamma^{\mu}_{\lambda\rho}+ (\beta_{(\lambda)}\delta_{\rho}^{\mu} + \beta_{(\rho)}\delta_{\lambda}^{\mu} -  \beta ^{(\mu)}g_{\lambda\rho}).
\end{equation}
Thus, the Ricci tensor $R_{\mu\nu}$, of the physical space-time, is given by
\begin{equation}
\label{physical Ricci tensor}
\hat{R}_{\mu\nu} = R_{\mu\nu} -(D_{(\mu)}\beta_{(\nu)} + D_{(\nu)}\beta_{(\mu)}) + 2 \beta_{(\mu)}\beta_{(\nu)} - g_{\mu\nu}[2\beta_{(\lambda)}\beta^{(\lambda)} + D_{(\lambda)}\beta^{(\lambda)}].
\end{equation}
Here $\beta_{(\alpha)} = \beta_{|(\alpha)}$; and $D$ is the operator of covariant differentiation constructed with the $\Gamma^{\mu}_{\lambda\rho}$ given above\footnote{The relation between $\hat{D}$ and $D$ is as follows: $\hat{D}_{(\mu)}V_{\nu} = D_{(\mu)}V_{\nu} - (\beta_{|(\mu)}V_{\nu} + \beta_{|(\nu)}V_{\mu} - \beta_{|(\alpha)}V^{\alpha}g_{\mu\nu})$.}. Therefore, $D_{(\lambda)}g_{\mu\nu} = 0$. In Eq. (\ref{physical Ricci tensor}) and in what follows, indexes are lowered and raised by the physical metric $g_{\mu\nu}$. 
\subsection{Four-dimensional energy-momentum tensor} 
We now proceed to construct the energy-momentum tensor $T_{\mu\nu}$ corresponding to the physical metric $g_{\mu\nu}$ specified in (\ref{physical metric}). Following the usual procedure, we define the energy-momentum tensor in $4D$ through the Einstein equations in $4D$, namely 
\begin{equation}
\label{def. of 4D matter}
T_{\mu\nu} = R_{\mu\nu} - \frac{1}{2}g_{\mu\nu}R,
\end{equation}
where $R = g^{\alpha\beta}R_{\alpha\beta}$ is the scalar curvature of the physical space-time. We also introduce the notation
\begin{equation}
\Psi_{\mu\nu} = \beta_{|(4)}g_{\mu\nu} + \frac{1}{2}g_{\mu\nu|(4)}.
\end{equation}
Such that $\hat{\Psi}_{\mu\nu} = e^{2\beta}\Psi_{\mu\nu}$; $\hat{\Psi} = \Psi = \Psi_{\mu\nu}g^{\mu\nu}$, and $\hat{\Psi}^{\mu\lambda} = e^{-2\beta}\Psi^{\mu\lambda}$.
 
Now, we  substitute Eq. (\ref{physical Ricci tensor}) into 
(\ref{Ricci tensor in the local basis}) and change the covariant derivatives from $\hat{D}$ to $D$. Then, using the field equations (\ref{field equations in the local frame}) and the above notation we obtain the scalar curvature of the physical space-time as follows 
\begin{equation}
\label{4D scalar curvature}
R =  \epsilon e^{2\beta}(\Psi^2 - \Psi_{\lambda\rho}\Psi^{\lambda\rho}) + \frac{3\epsilon}{4}e^{-2\beta}{\cal F}_{(\lambda)(\rho)}{\cal F}^{(\lambda)(\rho)} + 6(D_{(\mu)}\beta^{(\mu)} + \beta_{(\mu)}\beta^{(\mu)})- 2 k e^{2\beta}\hat{\cal T}^{(4)}_{(4)}. 
\end{equation}
In a similar way, we obtain from Eq. (\ref{def. of 4D matter}) an effective energy momentum tensor that clearly separates into six parts, namely,
\begin{equation}
\label{effective EMT from Einstein's equations}
R_{\mu\nu} -\frac{1}{2}g_{\mu\nu}R = T^{(I)}_{\mu\nu} + T^{(II)}_{\mu\nu} + T^{(III)}_{\mu\nu} + T^{(IV)}_{\mu\nu} + T^{(V)}_{\mu\nu} + T^{(VI)}_{\mu\nu},
\end{equation}
where
\begin{eqnarray}
\label{explicit EMT from Einstein's equations}
T^{(I)}_{\mu\nu}&=& \epsilon e^{2\beta}\left[\Psi_{\mu\nu|(4)} - 2\Psi_{\mu\rho}\Psi_{\nu}^{\rho} + (\Psi + 2 \beta_{|(4)})\Psi_{\mu\nu}
 - \frac{1}{2}g_{\mu\nu}(\Psi^2 + \Psi_{\lambda\rho}\Psi^{\lambda\rho} + 2 \Psi_{|(4)})\right],\nonumber \\
T^{(II)}_{\mu\nu}&=& \frac{1}{2}\left(D_{(\mu)}{\cal F}_{(\nu)} + D_{(\nu)}{\cal F}_{(\mu)}\right) + {\cal F}_{(\mu)}{\cal F}_{(\nu)} -
g_{\mu\nu}\left(D_{(\rho)}{\cal F}^{(\rho)} + {\cal F}_{(\rho)}{\cal F}^{(\rho)}\right),\nonumber \\
T^{(III)}_{\mu\nu}&=& \left(D_{(\mu)}{\beta}_{(\nu)} + D_{(\nu)}{\beta}_{(\mu)}\right) -2 {\beta}_{(\mu)}{\beta}_{(\nu)} - g_{\mu\nu}\left(2D_{(\rho)}{\beta}^{(\rho)} + {\beta}_{(\rho)}{\beta}^{(\rho)}\right),\nonumber \\
T^{(IV)}_{\mu\nu}&=& \epsilon\frac{e^{-2\beta}}{2}\left[{\cal F}_{(\mu)(\rho)}{\cal F}_{(\nu)}\;^{(\rho)} - \frac{1}{4} g_{\mu\nu}{\cal F}_{(\lambda)(\rho)}{\cal F}^{(\lambda)(\rho)}\right],\nonumber \\
T^{(V)}_{\mu\nu}&=& - \left(\beta_{(\mu)}{\cal F}_{(\nu)} + \beta_{(\nu)}{\cal F}_{(\mu)} + g_{\mu\nu}\beta_{(\alpha)}{\cal F}^{(\alpha)}\right),\nonumber \\
T^{(VI)}_{\mu\nu}&=& k \hat{e}_{(\mu)}^{A}\hat{e}^{B}_{(\nu)}{\cal T}_{AB}.
\end{eqnarray}
Let us note several features of the above equations:

1) $T^{(I)}_{\mu\nu}$ depends only on the derivatives along the  fifth dimension; $T^{(II)}_{\mu\nu}$ depends only on the four-vector ${\cal F}_{(\mu)}$; $T^{(III)}_{\mu\nu}$ depends only on the conformal factor; $T^{(IV)}$ depends only on the antisymmetric tensor ${\cal F}_{(\mu)(\nu)}$; $T^{(V)}_{\mu\nu}$ is a kind of ``interaction"  between the fields ${\cal F}_{(\mu)}$ and $\beta_{(\mu)}$; finally $T^{(VI)}_{\mu\nu}$ represents the space-time projection of the five-dimensional energy-momentum tensor on a general local basis.  

2) The interpretation of matter content of the space-time is fixed solely by the choice of four  space-time basis vectors $\hat{e}^{(\mu)}_{A}$. They in turn define $\hat{e}^{A}_{(4)}$ from (\ref{definition of psi}) and (\ref{extra basis vector}). Once this choice is made none of the contributions to $T_{\mu\nu}$ can be altered by changing coordinates in $5D$.

3) The factor $\epsilon$ appears in front of $T_{\mu\nu}^{(I)}$ and $T_{\mu\nu}^{(IV)}$. Therefore their effects on gravity in $4D$ will depend on whether the extra dimension is spacelike or timelike. The contribution from the other components of the energy-momentum tensor is the same in both cases. This can also be seen from (\ref{4D scalar curvature}). 

4) The special nature of $T_{\mu\nu}^{(I)}$ and $T_{\mu\nu}^{(IV)}$ (which we will see describe ordinary matter and electromagnetic field, respectively) is also reflected by the multiplicative factor in front of them. While the contribution from ordinary matter is enhanced by a factor $e^{2\beta}$, the one from the electromagnetic field is diminished in the same proportion. A similar effect occurs with the scalar curvature (\ref{4D scalar curvature}). 

5) Notice the absence of constants coefficients in (\ref{effective EMT from Einstein's equations}). The only ``arbitrary" assumption we have made so far is in the choice of the norm of vector $\psi$ in  (\ref{definition of psi}). Therefore, one could ask whether it is possible or not to introduce some particular constants in (\ref{effective EMT from Einstein's equations}) by changing this norm, say putting $\gamma_{AB}\psi^{A}\psi^{B} = \epsilon N^2$, $N = constant$. The answer to this question is negative; the whole construction is invariant under this change.

For completeness we provide an alternative expression, which is useful in particular applications. It is obtained from the field equations (\ref{field equations in the local frame}) and $\hat{\cal R}_{(4)(4)}$ in\footnote{Here we use that $\hat{\cal R}^{(4)}_{(4)} = (k/3)(2\hat{\cal T}^{(4)}_{(4)} - \hat{\cal T}^{(\alpha)}_{(\alpha)})$.} (\ref{Ricci tensor in the local basis}). 
\begin{eqnarray}
\label{wave equation}
D_{(\lambda)}{\cal F}^{(\lambda)} + {\cal F}_{(\lambda)}(2 \beta^{(\lambda)} + {\cal F}^{(\lambda)}) = \epsilon \frac{e^{-2\beta}}{4}{\cal F}_{(\lambda)(\rho)}{\cal F}^{(\lambda)(\rho)}\nonumber \\
 - e^{2\beta}\left(\epsilon(\Psi_{|(4)} + {\Psi}_{\lambda\rho}\Psi^{\lambda\rho}) + \frac{k}{3}(2 \hat{\cal T}^{(4)}_{(4)} - \hat{\cal T}^{(\lambda)}_{(\lambda)}) \right).
\end{eqnarray}
 
Finally, we stress the fact that  in Eqs. (\ref{physical Ricci tensor})-(\ref{wave equation}) we have lowered and raised indexes with the help of the physical metric $g_{\mu\nu}$.      
\subsection{Maxwell-like equations}
Equations (\ref{4D scalar curvature})-(\ref{effective EMT from Einstein's equations}) constitute eleven of the fifteen field equations in five dimensions. The other four are obtained from (\ref{field equations in the local frame}), (\ref{Ricci tensor in the local basis}) and (\ref{physical metric}), viz.,
\begin{equation}
\label{Maxwell Eq. with sources}
D_{(\lambda)}\left[\frac{e^{-2\beta}}{2}{\cal F}_{(\mu)}\;^{(\lambda)} + P_{\mu}^{\lambda}\right] + e^{-2\beta}{\cal F}_{(\mu)(\rho)}({\cal F}^{(\rho)} + \beta^{(\rho)}) + \epsilon \beta_{(\sigma)}(4\Psi_{\mu}^{\sigma} - \delta_{\mu}^{\sigma}\Psi) = k\hat{\cal T}_{(\mu)}^{(4)}.
\end{equation}
where
\begin{equation}
\label{conserved quantity in JPdeL}
P_{\mu}^{\lambda} = \epsilon (\hat{\Psi}_{\mu}^{\lambda} - \delta_{\mu}^{\lambda}\hat{\Psi}) = \epsilon (\Psi_{\mu}^{\lambda} - \delta_{\mu}^{\lambda}\Psi)
\end{equation}

We will see that the four equations (\ref{Maxwell Eq. with sources}) can be reduced to the inhomogeneous Maxwell equations, for electrical and magnetic field, containing sources. While the homogeneous Maxwell equations, ie., those without sources can be obtained  from the condition 
\begin{equation}
\label{condition on cal F}
{\cal F}_{(A)(B)|(C)} + {\cal F}_{(B)(C)|(A)} + {\cal F}_{(C)(A)|(B)} = {\cal F}_{(A)(B)}{\cal F}_{(C)(4)} + {\cal F}_{(B)(C)} {\cal F}_{(A)(4)} + {\cal F}_{(C)(A)}{\cal F}_{(B)(4)},
\end{equation}
which follows from the antisymmetric nature of ${\cal F}_{(A)(B)}$, and the commutativity relation (\ref{conmutativity relations}). This condition requires the identical fulfillment of    
  
\begin{equation}
\label{extra Maxwell equations}
{\cal F}_{(\mu)(\nu)|(4)} = {\cal F}_{(\mu)|(\nu)} - {\cal F}_{(\nu)|(\mu)},
\end{equation}
and 
\begin{equation}
\label{Gen. of source free Maxwell equations}
{\cal F}_{(\mu)(\nu)|(\lambda)} + {\cal F}_{(\nu)(\lambda)|(\mu)} + {\cal F}_{(\lambda)(\mu)|(\nu)} = {\cal F}_{(\mu)(\nu)}{\cal F}_{(\lambda)} + {\cal F}_{(\nu)(\lambda)} {\cal F}_{(\mu)} + {\cal F}_{(\lambda)(\mu)}{\cal F}_{(\nu)}.
\end{equation}
 It is clear that, this identity under certain conditions should lead to the one for the electromagnetic tensor.    

\section{Coordinate Frame}

The equations in Sec. 4 are rather general, they are useful for the interpretation of $5D$ metrics in an arbitrary frame. The choice of any particular local frame does not impose restrictions on the geometry. However, it should be clear that, the {\em physical} interpretation of any given metric in $5D$ crucially depends on this choice. In fact, the analysis of a given metric in two different (local) frames, will result in two different interpretations for the material content of the space-time. 

Specific results in the literature frequently assume that the first four coordinates correspond to the physical space-time, while the remaining one is the extra dimension, viz.,  

\begin{equation}
\label{special coordinates}
x^{\mu}= {\xi}^{\mu}.
\end{equation}
The space-time basis vectors are
\begin{eqnarray}
\label{space-time basis vectors}
\hat{e}^{(0)}_{A}&=& (1, 0, 0, 0, 0),\nonumber \\
\hat{e}^{(1)}_{A}&=& (0, 1, 0, 0, 0),\nonumber \\
\hat{e}^{(2)}_{A}&=& (0, 0, 1, 0, 0),\nonumber \\
\hat{e}^{(3)}_{A}&=& (0, 0, 0, 1, 0).
\end{eqnarray}
From (\ref{definition of psi}) and (\ref{extra basis vector}) we find
\begin{equation}
\label{e4}
\hat{e}^{A}_{(4)}= \epsilon{\psi^A}= (0, 0, 0, 0, \frac{\epsilon}{\Phi}),
\end{equation}
where we have set $\gamma_{44}= \epsilon \Phi^{2}$. The above set of vectors constitute what we call {\em coordinate frame}.

The associated basis vectors are given by (\ref{general relation between basis vectors}). Denoting $\gamma_{\mu 4}= \epsilon \Phi^2 A_{\mu}$, we obtain
\begin{eqnarray}
\label{associated basis vectors}
\hat{e}^{A}_{(0)}&=& (1, 0, 0, 0, -A_{0}),\nonumber \\ 
\hat{e}^{A}_{(1)}&=& (0, 1, 0, 0, -A_{1}),\nonumber \\ 
\hat{e}^{A}_{(2)}&=& (0, 0, 1, 0, -A_{2}),\nonumber \\ 
\hat{e}^{A}_{(3)}&=& (0, 0, 0, 1, -A_{3}),\nonumber \\ 
\hat{e}^{(4)}_{A}&=& \epsilon \Phi(A_{0}, A_{1}, A_{2}, A_{3}, 1).
\end{eqnarray}
The $5D$ line element takes the well known form  
\begin{eqnarray}
\label{special metric} 
d{\cal S}^2 &=& e^{2\beta}g_{\mu\nu}dx^{\mu}dx^{\nu}+ \epsilon \Phi^2\left(d{\xi}^4 + A_{\mu}dx^{\mu}\right)^2,\nonumber \\
\hat{g}_{\mu\nu}&=& e^{2\beta} g_{\mu\nu} = \gamma_{\mu\nu}-\epsilon\Phi^2A_{\mu}A_{\nu}.
\end{eqnarray}

We will keep the use of $\xi^{4}$, in order to avoid any confusion with the ``physical" displacement along the extra dimension\footnote{In this frame, space-time displacements are $dx^{\mu} = d\xi^{\mu}$, while the ones along the extra dimension are $dx^{(4)}= \hat{e}^{(4)}_{A}d{\xi^A}= \epsilon \Phi(d{\xi}^4 + A_{\mu}dx^{\mu})$.}.
From (\ref{antisymmetric tensor}), and using the above basis vectors, we obtain the specific form of ${\cal F}_{(\mu)(\nu)}$ and ${\cal F}_{(\mu)}$ in the coordinate frame. Namely,   
\begin{eqnarray}
\label{Gen of EM tensor in special coordinates}
{\cal F}_{(\mu)(\rho)}&=& \epsilon \Phi\left(A_{\rho|(\mu)}- A_{\mu|(\rho)}\right),\nonumber \\
{\cal F}_{(\mu)}&=& \frac{\Phi_{|(\mu)}}{\Phi}- \epsilon \Phi A_{\mu|(4)}.
\end{eqnarray}
In this frame the rules for derivation, for an arbitrary function $\phi$, are 
\begin{eqnarray}
\label{derivation rules}
\phi_{|(4)} &=& \epsilon \frac{\phi_{,4}}{\Phi},\nonumber \\
\phi_{|(\mu)} &=& \phi_{,\mu} - \phi_{,4}A_{\mu}, 
\end{eqnarray}
where a comma denotes partial differentiation. 
The above quantities are invariant under the set of ``gauge" transformations
\begin{eqnarray}
\label{Allowed transformations}
x^{\mu} &=& \bar{x}^{\mu},\nonumber \\ \xi^{4}&=& \bar{\xi}^{4}+ f(\bar{x}^{0},\bar{x}^{1},\bar{x}^{2},\bar{x}^{3}),
\end{eqnarray}
that keeps the shape of the line element (\ref{special metric}). As a matter of fact, all physical quantities in equations (\ref{4D scalar curvature})-(\ref{Gen. of source free Maxwell equations}) are invariant under these transformations\footnote{Under transformation (\ref{Allowed transformations}); $\bar{\gamma}_{\mu\nu} = \gamma_{\mu\nu} + \epsilon\Phi^{2}(A_{\mu}f_{,\nu} + A_{\nu}f_{,\mu} + f_{,\mu}f_{,\nu})$ and $\bar{A}_{\mu} = (A_{\mu} + f_{,\mu})$. The induced metric $\hat{g}_{\mu\nu}$ as well as the physical one $g_{\mu\nu}$ remain invariant. The derivatives change as $\bar{g}_{\mu\nu,\lambda} = g_{\mu\nu,\lambda} + g_{\mu\nu,4}f_{,\lambda}$, but the Christoffel symbols $\hat{\Gamma}^{\mu}_{\alpha\beta}$ and $\Gamma^{\mu}_{\alpha\beta}$ remain invariant. Also, 
$\bar{\Phi}_{,\mu}=\Phi_{,\mu}+ \Phi_{,4}f_{,\mu}$, $\bar{A}_{\mu,\nu} = A_{\mu,\nu} + A_{\mu,4}f_{,\nu} + f_{,\mu,\nu}$ and $\bar{A}_{\mu|(\nu)} = A_{\mu|(\nu)} + f_{,\mu,\nu}$. Therefore, the antisymmetric tensor ${\cal F}_{(\mu)(\rho)}$ and the four-vector ${\cal F}_{(\mu)}$ remain invariant.}. 

In what follows we will use the quantity
\begin{eqnarray}
\label{hatF}
\hat{F}_{\mu\rho}&=& \left(A_{\rho|(\mu)}-A_{\mu|(\rho)}\right) = \left(D_{(\mu)}A_{\rho}-D_{(\rho)}A_{\mu}\right)\nonumber \\
&=& \left(A_{\rho,\mu}-A_{\mu,\rho}\right) + \left(A_{\rho}A_{\mu,4}-A_{\mu}A_{\rho,4}\right).
\end{eqnarray}
Which, in the case of no dependence on the extra variable, reduces to the usual electromagnetic tensor, viz., 
\begin{equation}
\label{usual EMT}
{F}_{\mu\rho}= (A_{\rho,\mu}-A_{\mu,\rho}).
\end{equation}
Therefore we will refer to (\ref{hatF}) as the {\it generalized electromagnetic tensor}. Thus, in what follows
\begin{equation}
\label{cal F in terms of F}
{\cal F}_{(\mu)(\nu)} = \epsilon \Phi \hat{F}_{\mu\nu}
\end{equation}

Since the choice of the coordinate-frame seems to be the simplest one, for a given metric, the question may arise of whether a simplified version for the effective energy-momentum-tensor (\ref{effective EMT from Einstein's equations})-(\ref{explicit EMT from Einstein's equations}) can be derived in a simpler way by employing the coordinate frame from the outset.

The answer to this question is negative, at least for the general case where cylindricity is not assumed and $\gamma_{4\mu} \neq 0$. Indeed, the explicit coordinate-frame version of (\ref{explicit EMT from Einstein's equations}) is quite cumbersome. 

\subsection{$\xi^{4}$-dependence of induced four-dimensional fields}

Since the bulk metric $\gamma_{AB}$ is allowed to be a function of the extra coordinate, the induced physical fields $g_{\mu\nu}$, $\Phi$ and $\hat{F}_{\mu\nu}$ depend on the space-time coordinates and also on the fifth coordinate. However the $\xi^{4}$-dependence of these fields is non dynamic. Besides, these fields are gauge invariant under (\ref{Allowed transformations}).   

The physical meaning of (\ref{Allowed transformations}) is that $\xi^{4}$ can take any value independently at each space-time point. It reflects the arbitrariness in the choice of origin for $\xi^{4}$ at each space-time point. Therefore, without loss of generality we can set $\xi^{4} = \xi^{4}_{0} = const$, everywhere in space-time. Consequently, $4D$ quantities will depend on $x^{0}$, $x^{1}$, $x^{2}$ and $x^{3}$ only, but {\em not} on $\xi^{4}$.

\section{Particular Cases and Previous Results}

In general, we cannot expect to obtain simplified versions for (\ref{effective EMT from Einstein's equations})-(\ref{explicit EMT from Einstein's equations}) except in the case where we introduce some physical simplifying assumptions. The simplest versions are generated by (\ref{conmutativity relations}). They correspond to the cases where the (directional) derivatives commute. They are: 

(i) no dependence of the extra variable, 

(ii) no electromagnetic field, 

(iii) no dependence of the extra variable and no electromagnetic field.      

\subsection{Classical Kaluza-Klein theory: no dependence on the extra coordinate and $\gamma_{\mu 4} \neq 0$}

In this case $T^{(I)}_{\mu\nu}$ from (\ref{effective EMT from Einstein's equations}) vanishes. The other components reduce to
\begin{eqnarray}
\label{First particular case}
T^{(II)}_{\mu\nu} &=& \frac{1}{\Phi}(\Phi_{\mu;\nu} - g_{\mu\nu}\Phi^{\alpha}_{;\alpha}),\nonumber \\
T^{(III)}_{\mu\nu} &=& 2\beta_{\mu;\nu} - 2\beta_{\mu}\beta_{\nu} - g_{\mu\nu}(2\beta^{\alpha}_{;\alpha} + \beta_{\alpha}\beta^{\alpha}),\nonumber \\
T^{(IV)}_{\mu\nu} &=& \epsilon \frac{\Phi^2 e^{-2\beta}}{2}(F_{\mu\rho}F_{\nu}\;^{\rho} - \frac{1}{4}g_{\mu\nu}F_{\lambda\rho}F^{\lambda\rho}),\nonumber \\
T^{(V)}_{\mu\nu} &=& - \frac{1}{\Phi}\left(\beta_{\mu}\Phi_{\nu} +  \beta_{\nu} \Phi_{\mu} + g_{\mu\nu}\beta_{\alpha}\Phi^{\alpha}\right),
\end{eqnarray}
where semicolon denotes covariant differentiation and $\phi_{\alpha} = \phi_{,\alpha}$. 

We now go back to our comment (5), at the end of Sec. 4.2. In order to ``derive" the dimensional units in $4D$ from $5D$, some authors put a multiplicative coefficient in front of the extra part of the line element (\ref{special metric}). The change $\Phi \rightarrow K \Phi$ leads to a coefficient $K^2$ in front of $T^{(IV)}_{\mu\nu}$, while the other components remain unchanged.

Equation (\ref{wave equation}) reduces to the wave equation for the scalar field,
\begin{equation}
\Phi^{\alpha}_{;\alpha} + 2 \Phi_{\alpha}\beta^{\alpha} = \epsilon \frac{e^{-2\beta}}{4}\Phi^3 F_{\lambda\rho}F^{\lambda\rho} - \Phi e^{2\beta}\hat{\cal R}^{(4)}_{(4)},
\end{equation}
while equations (\ref{Maxwell Eq. with sources}) and (\ref{Gen. of source free Maxwell equations}) yield  equations of ``electromagnetic type"\footnote{In (\ref{Maxwell eq. part. case}) 
we have used $\hat{\cal T}^{(\mu)(4)} = \hat{e}^{(\mu)A}\hat{e}^{(4)B}{\cal T}_{AB} = \epsilon\hat{e}^{(\mu)A}\hat{e}_{(4)}^{B}{\cal T}_{AB} = {\cal T}^{(\mu)}_{4}/\Phi$. Basis indexes are lowered and raised with the induced metric.}

\begin{equation}
\label{Maxwell eq. part. case}
\left(\Phi^3 F^{\mu\lambda}\right)_{;\lambda} = 2\epsilon \Phi e^{4\beta}k{\cal T}^{(\mu)}_{4}
\end{equation}

\begin{equation}
F_{[\mu\nu;\lambda]} = 0.
\end{equation}

\subsubsection{$\beta = 0$}
In this case we recover well known results of classical Kaluza-Klein theory. Setting $\Phi^2 \rightarrow (4G/c^4)\Phi^2$ and $k = (8\pi G/c^4)$, we obtain $T^{(III)}_{\mu\nu} = 0$, $T^{(V)}_{\mu\nu} = 0$ and $(\Phi^3 F^{\mu\lambda})_{;\lambda} = (4\pi/c)J^{\mu}$, where $J^{\mu} = \epsilon c\Phi{\cal T}^{(\mu)}_{4}$. In this case, our results are {\em identical} to those in Ref. \cite{Nodvik}. In this reference the connection between ${\cal T}_{AB}$ and the four-vector charge-current density $J^{\mu}$ is taken as a {\it postulate}. In our approach this connection is {\it not} a postulate but is a byproduct of the dimensional reduction.

\subsubsection{$\beta \neq 0$}
 We recover other particular cases in the literature. See bellow in Sec. 6.3.

 \subsection{STM and brane-world theory: explicit dependence on the extra variable and $\gamma_{\mu 4} = 0$}

The condition $\gamma_{\mu 4} = 0$ requires $A_{\mu} = 0$. The components of the energy-momentum tensor are identical to (\ref{First particular case}), except for the fact that now $T^{(IV)}_{\mu\nu} = 0$.
In this case $T^{(I)}_{\mu\nu}$ is the combination of two parts\footnote{$^{(\varphi)}T_{\mu\nu}^{(I)} = \epsilon\frac{e^{2\beta}}{\Phi^2}\left[\stackrel{\ast}{\varphi}_{\mu\nu} -\frac{\stackrel{\ast}{\Phi}}{\Phi}\varphi_{\mu\nu} -2 \varphi_{\mu\rho}\varphi^{\rho}_{\nu} + \varphi\varphi_{\mu\nu} + g_{\mu\nu}(\frac{\stackrel{\ast}{\Phi}}{\Phi}\varphi - \stackrel{\ast}{\varphi} - \frac{1}{2}\varphi^2 - \frac{1}{2}\varphi_{\lambda\rho}\varphi^{\lambda\rho})\right]$ and
$^{(\beta)}T^{(I)}_{\mu\nu} = \epsilon \frac{e^{2\beta}}{\Phi^2}\left[ \stackrel{\ast}{\beta}(4 \varphi_{\mu\nu} - 4 \varphi g_{\mu\nu} + 3 \frac{\stackrel{\ast}{\Phi}}{\Phi}g_{\mu\nu}) - 3g_{\mu\nu}(\stackrel{\ast \ast}{\beta} + 2 (\stackrel{\ast}{\beta})^2)\right]$.}
. One of them, $^{(\varphi)}T^{(I)}_{\mu\nu}$, depends only on the extra derivatives of the physical metric. The other part, $^{(\beta)}T^{(I)}_{\mu\nu}$, contains the extra derivatives of $\beta$. 

In what follows an asterisk denotes partial differentiation with respect to the extra coordinate, and  
\[
\varphi_{\mu\nu} = \frac{1}{2}\stackrel{\ast}{g}_{\mu\nu},\;\;\;\varphi^{\mu\nu} = - \frac{1}{2}(g^{\mu\nu})^{\ast},\;\; \varphi = g^{\mu\nu}\varphi_{\mu\nu}.\] 

The scalar field satisfies the equation
\begin{equation}
\label{equation for phi in STM}
\Phi^{\alpha}_{;\alpha} = - 2 \Phi_{\alpha}\beta^{\alpha} - \epsilon \frac{e^{2\beta}}{\Phi}\left[(\stackrel{\ast}{\varphi} - \frac{\stackrel{\ast}{\Phi}}{\Phi}\varphi + \varphi_{\lambda\rho}\varphi^{\lambda\rho}) + 4(\stackrel{\ast \ast}{\beta} + (\stackrel{\ast}{\beta})^2 - \frac{\stackrel{\ast}{\Phi}}{\Phi}\stackrel{\ast}{\beta} + \frac{1}{2}\stackrel{\ast}{\beta}\varphi ) + \epsilon {\Phi}^{2} \hat{\cal R}^{(4)}_{(4)}\right].
\end{equation}
Using this expression we obtain,
\begin{eqnarray}
\label{T1 + T2}
& & T_{\mu\nu}^{(I)} + T_{\mu\nu}^{(II)} = \left[\frac{\Phi_{\mu;\nu}}{\Phi} + 2g_{\mu\nu}\frac{\Phi_{\lambda}}{\Phi}\beta^{\lambda}\right]\nonumber \\
 &+& \frac{\epsilon e^{2\beta}}{\Phi^2}\left[(\stackrel{\ast}{\varphi}_{\mu\nu} - \frac{\stackrel{\ast}{\Phi}}{\Phi}\varphi_{\mu\nu} - 2\varphi_{\mu\rho}\varphi^{\rho}_{\nu} + \varphi\varphi_{\mu\nu}) - \frac{1}{2}g_{\mu\nu}(\varphi^2 - \varphi_{\lambda\rho}\varphi^{\lambda\rho})\right]\nonumber \\
&+& \frac{\epsilon e^{2\beta}}{\Phi^2}\left[ 4 \stackrel{\ast}{\beta}\varphi_{\mu\nu} - g_{\mu\nu}\left( 2(\stackrel{\ast}{\beta})^2 + \frac{\stackrel{\ast}{\Phi}}{\Phi}\stackrel{\ast}{\beta} + 2 \varphi\stackrel{\ast}{\beta} - \stackrel{\ast \ast}{\beta}  -  \epsilon {\Phi}^{2}\hat{\cal R}^{(4)}_{(4)}\right)\right].
\end{eqnarray}
These four-dimensional quantities have to be evaluated at some $\xi^{4} = const$. In brane-world models, for convenience, $\xi^{4} = 0$ is usually chosen.

\subsubsection{$\beta = 0$}

Here we recover the effective energy-momentum tensor in STM and brane-world models.

 In the context of STM, it is straightforward to verify that our equations (with $\hat{\cal R}^{(4)}_{(4)} = 0$) become identical to those in Ref.\cite{Wesson and JPdeL}. From (\ref{Maxwell Eq. with sources}) it follows that the $4D$ tensor $P_{\mu\nu}$ is a conserved quantity provided $\hat{\cal T}^{(4)}_{(\mu)} = 0$, viz., $D_{\lambda}P^{\lambda}_{\mu} = 0$ \cite{Wesson JPdeL Conserved quantity}. These equations form the basis of STM. From a four-dimensional point of view, the empty $5D$ equations  look as the Einstein equations with (effective) matter. In the case where the {\em bulk} metric is independent of the extra dimension, equations (\ref{equation for phi in STM}) and (\ref{T1 + T2}) show  that the effective $4D$  energy-momentum tensor is traceless, $T^{\mu}_{(eff) \mu} = 0$. In other words, independence of the $5D$ metric from the extra coordinate implies a radiation-like equation of state. Thus the existence of other forms of $4D$ matter crucially relies on the $\xi^{4}$-dependence of the bulk metric. A detailed investigation shows that we can recover a number of  equations of state commonly used in astrophysics and cosmology \cite{JPdeL 1}, \cite{JPdeL 2}.

In the brane-world scenario, $\varphi_{\mu\nu}$ is proportional to the extrinsic curvature of the brane, which in turn is related, through Israel's boundary conditions, to the energy-momentum tensor of ordinary matter on the brane. The quantities   $ \stackrel{\ast}{\varphi}_{\mu\nu}$, that is the second derivatives of the metric with respect to the extra coordinate, as well as $\stackrel{\ast}{\Phi}$ are related to the nonlocal Weyl corrections from the free gravitational field in the bulk. With this interpretation, one can show that  (\ref{equation for phi in STM}) and (\ref{T1 + T2}) look exactly  as the equations for gravity in brane-world models (for details see Ref. \cite{STM-Brane}).  
 
\subsubsection{$\beta \neq 0$}  

Let us first consider the ``canonical metric" \cite{Wesson book}. This metric has $e^{2\beta} = (l/L)^2$, where $l$ is the extra coordinate and $L$ is a dimensional coefficient. The (physical) space-time metric is allowed to depend on the extra variable. With this choice $T^{(III)}_{\mu\nu} = T^{(IV)}_{\mu\nu} = T^{(V)}_{\mu\nu} = 0$. In this case, Eq. (\ref{T1 + T2}) correctly reproduces (with $\epsilon = -1$) the effective energy-momentum tensor as discussed in Ref. \cite{Wesson book} (pages 157, 159).

Finally, we consider the so called ``warped metrics". These are similar to the canonical metrics. The difference is that here $\Phi = 1$ and $\beta$ is an arbitrary function of the extra variable, and the space-time metric is independent on it. In this case 
\begin{equation}
T_{\mu\nu}^{(I)} = -3 \epsilon g_{\mu\nu}e^{2\beta}(\stackrel{\ast \ast}{\beta} + 2\stackrel{\ast}{\beta}^2).
\end{equation}
If we set $e^{2\beta} = \Omega$ it reduces to the one obtained in Ref. \cite{Seahra and Wesson}, with $\epsilon = -1$.

\subsection{No dependence on the extra coordinate and $\gamma_{\mu 4} = 0$}

We now consider the case where there is no electromagnetic field and that the conformal factor $e^{2\beta}$ is some arbitrary function of the scalar field, say, $e^{2\beta} = F(\Phi)$. From (\ref{First particular case}) we obtain $T_{\mu\nu}^{(IV)} = 0$ and 
\begin{eqnarray}
T_{\mu\nu}^{(II)} + T_{\mu\nu}^{(III)} + T_{\mu\nu}^{(V)} = (1 + 2\Phi f)\left(\frac{\Phi_{\mu}}{\Phi}\right)_{;\nu} + (1 + 2\Phi^2f_{\Phi} -2\Phi^2f^2)\frac{\Phi_{\mu}\Phi_{\nu}}{\Phi^2}\nonumber \\
 - g_{\mu\nu}\left[ (1 + 2\Phi f))\left(\frac{\Phi_{\alpha}}{\Phi}\right)_{;\alpha} + (1 + 3\Phi f + 2\Phi^2 f_{\Phi} + \Phi^2 f^2)\frac{\Phi_{\alpha}\Phi^{\alpha}}{\Phi^2}\right],
\end{eqnarray}
where $f = (1/2F)(dF/d\Phi)$ and $f_{\Phi} = df/d\Phi$. 

If we now choose $F = \Phi^{2n}$, where $n$ is an arbitrary real constant, then we obtain the same results as in Refs. \cite{Kokarev 1}, and \cite{Kokarev 2} (See also Ref. \cite{Wesson book}, pag. 194).

\section{General Case: No Cylindricity and $\gamma_{4\mu} \neq 0$}

The conclusion from Sec. 6 is that our general equations (\ref{4D scalar curvature})-(\ref{Gen. of source free Maxwell equations}) reproduce well known results in the literature. This allows us to infer that (\ref{4D scalar curvature})-(\ref{Gen. of source free Maxwell equations}) are reliable and can be used in the general case for further investigations in gauge invariant Kaluza-Klein theory.

The general case, with no cylindricity and $\gamma_{4 \nu} \neq 0$, presents two new important features. 
 The first one is the anholonomic nature of the theory. In any local frame, including the coordinate one, the derivatives (\ref{conmutativity relations}) do not commute. The second new feature concerns the electromagnetic tensor. In order to keep the gauge invariance of the fields we are forced to consider (\ref{hatF}), and not (\ref{usual EMT}) as the physical relevant quantity\footnote{Under gauge transformations ${F}_{\mu\nu} \rightarrow F_{\mu\nu} + \left(A_{\nu,4}f_{\mu}- A_{\mu,4}f_{,\nu}\right)$, while $\hat{F}_{\mu\nu}$ remains invariant.}. We now proceed to show that, as a consequence of these new features, one can construct two conserved four-dimensional ``currents".  

\subsection{Induced effective ``magnetic charge" and ``magnetic current"}
As a direct consequence of these new features, equation (\ref{Gen. of source free Maxwell equations}) leads to ones that resemble Maxwell equations with magnetic charge and current. In order to show this, we substitute (\ref{cal F in terms of F}) into (\ref{Gen. of source free Maxwell equations}) and use (\ref{Gen of EM tensor in special coordinates}). We obtain

 \begin{equation}
\label{magnetic Maxwell equations}
\hat{F}_{\mu\nu|(\lambda)} + \hat{F}_{\nu\lambda|(\mu)} + \hat{F}_{\lambda\mu|(\nu)} = - (\hat{F}_{\mu\nu}{M}_{\lambda} + \hat{F}_{\nu\lambda}{M}_{\mu} + \hat{F}_{\lambda\mu}{M}_{\nu}).
\end{equation}
Here, and in what follows $M_{\nu} = A_{\nu,4} = \epsilon \Phi A_{\nu|(4)} = \stackrel{\ast}{A}_{\nu}$. Let us now introduce the quantities \cite{Landau and Lifshitz}  (Latin indexes run over 123)  
\begin{eqnarray}
\label{Invariant definition of E and B}
E_{i}  &=&  \hat{F}_{0i} \nonumber \\
B_{ij} &=& \hat{F}_{ij},
\end{eqnarray}
and define the vector ${\bf B}$ dual\footnote{With these definitions, the fields ${\bf E}$ and ${\bf B}$ are gauge invariant} to $B_{ij}$
\begin{equation}
B^{i} = - \frac{1}{2\sqrt{\lambda}}\epsilon^{ijk}B_{jk},
\end{equation}
where ${\lambda}$ is the determinant of the metric of the three-dimensional space\footnote{The space-time metric can be put in the form $ds^2 = [\sqrt{g_{00}}dx^0 +  (g_{0i}/\sqrt{g_{00}})dx^i]^2 + [-(g_{0j}g_{0k}/g_{00}) + g_{ik}]dx^idx^k$.  The metric of the three-dimensional space is $\lambda_{ij} = - g_{ij} + (g_{0j}g_{0k}/g_{00})$.}. Setting $\mu \rightarrow i$, $\nu \rightarrow j$ and $\lambda \rightarrow 0$ in (\ref{magnetic Maxwell equations}) and using above definitions\footnote{We note that $- g = g_{00}\lambda$, $B_{ij} = - \sqrt{\lambda}\epsilon_{ijk}B^k$  and  $({\bf a}\times {\bf b})^i = (1/2\sqrt{\lambda})\epsilon^{ijk}(a_{j}b_{k} - a_{k}b_{j})$. In the local frame, we use the following definitions for curl and divergence, respectively: $(\hat{ \bf \bigtriangledown}\times {\bf a})^i = (1/2\sqrt{\lambda})\epsilon^{ijk}(a_{k|(j)} - a_{j|(k)})$, and $\hat{ \bf \bigtriangledown}\cdot{\bf a} = (1/\sqrt{\lambda})(\sqrt{\lambda}a^{i})_{|(i)}$.} we get
\begin{eqnarray}
\label{rotE}
\hat{ \bf \bigtriangledown}\times {\bf E} &=& -\frac{1}{\sqrt{\lambda}}\frac{\partial(\sqrt{\lambda}{\bf B})}{\partial{x^{(0)}}} - \hat{\bf j}_{m},\nonumber \\
\hat{\bf j}_{m} &=& {M}_{0}{\bf B} - ({\bf E} \times\ {{\bf M}}). 
\end{eqnarray}
where ${\bf M}= ({M}_{1}, {M}_{2}, {M}_{3})$. We note that ${M}_{0}$ and ${\bf M}$ behave, respectively, as three-dimensional scalar and vector quantities with respect to $3D$ spatial transformations ${x}^i = x^i(\bar x^j)$, and are invariant under gauge transformations (\ref{Allowed transformations}). Now setting $\mu\rightarrow1$, $\nu\rightarrow2$, $\lambda\rightarrow3$ in (\ref{magnetic Maxwell equations}) we get
\begin{eqnarray}
\label{divB}
\hat{ \bf \bigtriangledown}\cdot {\bf B} &=& \hat{\rho}_{m},\nonumber \\
\hat{\rho}_{m} &=& - ({\bf B}\cdot {\bf M}).
\end{eqnarray}
We note the formal analogy between (\ref{rotE}), (\ref{divB}) and the Maxwell equations in the electric field ${\bf E}$ and magnetic field ${\bf B}$. The quantities $\hat{\rho}_{m}$ and $\hat{\bf j}_{m}$ play the role of ``magnetic charge" and ``magnetic current" density (in the local frame), respectively. We will refer to them as $induced$ or $effective$ magnetic charge and current. They satisfy the following equation
\begin{equation}
\label{conservation eq. in local frame}
\frac{1}{\sqrt{\lambda}}\frac{\partial}{\partial{x^{(0)}}}({\sqrt{\lambda}\hat{\rho}_{m}}) + \hat{ \bf \bigtriangledown}\cdot \hat{\bf j}_{m} =  {\bf E}\cdot (\hat{ \bf \bigtriangledown}\times {\bf M}),
\end{equation}
where we have used (\ref{extra Maxwell equations}), which now reads ${\hat{F}}_{\mu\nu,4} = (M_{\nu|(\mu)} - M_{\mu|(\nu)})$. This equation resembles the continuity equation, except for the appearance of the term on the right-hand-side. This term, is a direct consequence of the anholonomic nature of the theory.

In order to obtain the expression for the conserved $4D$ magnetic current we introduce the third-rank antisymmetric tensor
\begin{equation}
J_{\mu\nu\lambda} = (\hat{F}_{\mu\nu}M_{\lambda} - \stackrel{\ast}{\hat{F}}_{\mu\nu}A_{\lambda}) + (\hat{F}_{\nu\lambda}M_{\mu} - \stackrel{\ast}{\hat{F}}_{\nu\lambda}A_{\mu}) + (\hat{F}_{\lambda\mu}M_{\nu} - \stackrel{\ast}{\hat{F}}_{\lambda\mu}A_{\nu}),
\end{equation}
and define the four-vector $J^{\mu}_{m}$ dual to $J_{\lambda\rho\sigma}$
\begin{equation}
\label{magnetic current}
J^{\mu}_{m}= \frac{1}{3!\sqrt{- g}}\epsilon^{\mu\lambda\rho\sigma}J_{\lambda\rho\sigma}.
\end{equation}
From (\ref{magnetic Maxwell equations}) it follows that
\begin{equation}
\label{relation between magnetic current and EMT}
\sqrt{- g}J^{\rho}_{m} = - \frac{1}{2} \epsilon^{\rho\lambda\mu\nu}\hat{F}_{\mu\nu,\lambda}
\end{equation}
Now, by virtue of the commutativity of partial derivatives we obtain
\begin{equation}
\label{conservation of magnetic current}
\frac{1}{\sqrt{- g}}\frac{\partial}{\partial x^\mu}(\sqrt{- g}J^{\mu}_{m})= 0.
\end{equation}
This is the continuity equation and shows that the four-vector of ``effective" magnetic charge and current (\ref{magnetic current}) is a four-dimensional conserved quantity. 

In three-dimensional form the above equations read
\begin{equation}
\label{M1}
{ \bf \bigtriangledown}\times {\bf E} = -\frac{1}{\sqrt{\lambda}}\frac{\partial(\sqrt{\lambda}{\bf B})}{\partial t} - {\bf j}_{m}
\end{equation}
\begin{equation}
\label{source for the magnetic field}
 { \bf \bigtriangledown}\cdot {\bf B} = \rho_{m},
\end{equation}
where $\rho_{m} = \sqrt{g_{00}} J_{m}^{0}$, $j^{k}_{m} = \sqrt{g_{00}}J_{m}^{k}$ and the curl and divergence are defined as usual\footnote{$({ \bf \bigtriangledown}\times {\bf a})^i = (1/2\sqrt{\lambda})\epsilon^{ijk}(a_{k,j} - a_{j,k})$, and ${ \bf \bigtriangledown}\cdot{\bf a} = (1/\sqrt{\lambda})(\sqrt{\lambda}a^{i})_{,i}$. We note that $\hat{ \bf \bigtriangledown}\times {\bf a} = { \bf \bigtriangledown}\times {\bf a} + \stackrel{\ast}{\bf a}\times{\bf A}$ and $\hat{ \bf \bigtriangledown}\cdot{\bf a} = { \bf \bigtriangledown}\cdot{\bf a} - (1/\sqrt{\lambda})(\sqrt{\lambda}a^{i})^{\ast}{A_{i}}$. }.

Let us note several features of $J_{m}^{\mu}$:

1) Under  gauge transformations (\ref{Allowed transformations}) it transforms as 
\[
{\bar{J}}^{\rho}_{m} = J^{\rho}_{m} - \frac{1}{2\sqrt{-g}}\epsilon^{\rho\mu\nu\lambda}\stackrel{\ast}{\hat{F}}_{\mu\nu}f_{\lambda}.\]
However, this non-invariance is not a problem at all because (\ref{relation between magnetic current and EMT}) as well as the physical fields ${\bf E}$ and ${\bf B}$ are, by their definition (\ref{Invariant definition of E and B}), invariant under such transformations\footnote{We recall that under gauge transformation $\hat{\bar{F}}_{\alpha\beta, \mu} = \hat{F}_{\alpha\beta,\mu} + \stackrel{\ast}{\hat{F}}_{\alpha\beta}f_{\mu}$. ($\hat{F}_{\alpha\beta|(\mu)}$ is invariant). Therefore, the transformed quantities $\bar{J}^{\mu}_{m}$ and $\hat{\bar F}_{\mu\nu}$ satisfy $\sqrt{-g}\bar{J}^{\rho}_{m} = -  (\epsilon^{\rho\lambda\mu\nu}/2)\hat{\bar F}_{\mu\nu,\lambda}$ which is Eq. (\ref{relation between magnetic current and EMT}). This assures that the conservation of $J^{\mu}_{m}$ is gauge invariant. }.

2) The factor $\epsilon$ appears nowhere in the definition of $J_{m}^{\mu}$. This means that the induced magnetic current is the same regardless of whether the extra dimension is spacelike or timelike. Also, it is invariant under scale changes of the scalar field, viz., $\Phi\rightarrow K \Phi$.

3) The no-conservation of the magnetic charge $\hat{\rho}_{m}$ and current $\hat{\bf j}_{m}$, defined in the local frame, versus the conservation of ${\rho}_{m}$ and ${\bf j}_{m}$ defined above, is not surprising. The reason is that $\hat{\rho}_{m}$ is not the total induced magnetic charge. The total charge\footnote{A similar expression can be found for ${\bf j}_{m}$, namely, ${\bf j}_{m} = M_{0}{\bf B} - ({\bf E}\times {\bf M}) + (\stackrel{\ast}{\bf E}\times {\bf A}) + A_{0}[(\bf \bigtriangledown \times {\bf M}) + (\stackrel{\ast}{\bf M}\times {\bf A})]$.} (which is conserved) is given by ${\rho}_{m}$ and it is ${\rho}_{m} = -({\bf B}\cdot{\bf M}) - (\bf \bigtriangledown \times {\bf M})\cdot {\bf A}$. 

The conclusion from the above discussion is that $if$ the antisymmetric tensor $\hat{F}_{\mu\nu}$ can be identified with the electromagnetic one, then the theory predicts the existence of (classical) magnetic charge and current. In the case of  no dependence on the extra coordinate (\ref{source for the magnetic field}) reduces to the usual ${ \bf \bigtriangledown}\cdot {\bf B} = 0$, which allows the existence of static, point-like, magnetic monopoles with ${\bf B} = constant ({\bf r}/r^3)$ \cite{Gross and Perry}, \cite{Sorkin}.   

\subsection{Induced effective ``electric charge" and ``electric current"}
We now turn to the study of (\ref{Maxwell Eq. with sources}). After some manipulations, it can be written in as 
\begin{equation}
\label{Maxwell-like with physical sources}
\frac{1}{\sqrt{-g}}\frac{\partial}{\partial x^{\rho}}(\sqrt{- g}\Phi^3 \hat{F}^{\mu\rho}) = \frac{4\pi}{c}(J^{\mu} + J^{\mu}_{e,ind}),
\end{equation}
where we used the same identification between $\hat{\cal T}_{AB}$ and the four-vector charge and current density as in (\ref{Maxwell eq. part. case}) and Ref. \cite{Nodvik}. $(\partial/\partial {x^i})$ denotes usual partial derivative and $J^{\mu}_{e,ind}$ represents the induced current of geometrical origin, namely, 
\begin{eqnarray}
\label{induced electrical charge and current}
\frac{4\pi}{c} J^{\mu}_{e,ind}= & &\left[\frac{(\sqrt{- g}\Phi^3\hat{F}^{\mu\rho})^{\ast}}{\sqrt{- g}}A_{\rho}+ 2\Phi^3\hat{F}^{\mu\rho}M_{\rho}\right]\nonumber \\
   & &- 2\Phi^2e^{2\beta}\left[ \beta_{\sigma}(4\Psi^{\mu\sigma} - g^{\mu\sigma}\Psi) + \epsilon D_{(\sigma)}P^{\sigma\mu}\right].
\end{eqnarray}

Once again, by virtue of the commutativity of partial derivatives and the antisymmetric nature of $\hat{F}_{\mu\nu}$, the total current is a four-dimensional conserved quantity, viz., 
\begin{equation}
\label{conservation of electrical current}
\frac{1}{\sqrt{- g}}\frac{\partial}{\partial{x^{\mu}}}[\sqrt{- g }(J^{\mu}+J^{\mu}_{e,ind})] = 0
\end{equation}
The factor $\Phi^3$ in the left-hand-side of (\ref{Maxwell-like with physical sources}) can be interpreted in terms of classical vacuum polarization as in Ref. \cite{Nodvik}. Alternatively, the charge and current associated with this polarization, which is $-3(\Phi_{\rho}/\Phi)\hat{F}^{\mu\rho}$, can be put in the right hand-side of (\ref{Maxwell-like with physical sources}) as a part of the induced source.

Thus, in the general case (no cylindricity and $\gamma_{4\mu} \neq 0$), the dimensional reduction provides the set of Maxwell-like equations (\ref{relation between magnetic current and EMT}) (or (\ref{M1})-(\ref{source for the magnetic field}
) in three-dimensional form)  and (\ref{Maxwell-like with physical sources}).  The effective $4D$ sources, (\ref{magnetic current}) and (\ref{induced electrical charge and current}),  satisfy the usual conservation equations (\ref{conservation of magnetic current}) and (\ref{conservation of electrical current}). We recall that all these four-dimensional equations have to be evaluated at some $\xi^{4} = const$.

\section{Conclusions}
The splitting procedure we have used here is not intended to solve the field equations. Rather, it is a method to construct the appropriate four-dimensional quantities that are invariant under coordinate transformations in $5D$. It has a number of advantages:

(i) The whole discussion is independent of the specific coordinates used in the five-dimensional manifold and the way we choose coordinates in $4D$.

(ii) It gives a unified approach to a number of particular scenarios discussed in the literature. 

(iii) It uncovers the intrinsic anholonomic nature of the theory where, from one hand, the metric is allowed to depend on the extra coordinates and, the other hand, gauge invariance is required. 

(iv) It provides some new physical insight. In particular, it shows how to generalize the electromagnetic tensor  in such a way that the corresponding physical fields be gauge invariant. It also allows us to speculate about the possible relation between the conserved quantity $J^{\mu}_{m}$ and the existence of classical magnetic charge. Since it depends solely on the variations of the metric along the extra dimension, this would be a unique effect from the extra dimension.

(v) Since the discussion is independent of the number of dimensions, our results can be used in brane world theories to interpret physical quantities in one lower dimension.

The picture that comes out from our paper is the following. For any given five-dimensional metric, in arbitrary coordinates in $5D$, we choose our space-time basis vectors as we wish. Then  from (\ref{definition of psi}) and (\ref{extra basis vector}) we find $\hat{e}_{(4)}^{A}$. This completes our local frame. Next we construct the induced $4D$ metric as in (\ref{4D metric}) as well as the quantities ${\cal F}_{\mu\nu}$, ${\cal F}_{\mu}$ and $\hat{\Psi}_{\mu\nu}$. 

Next we proceed to calculate the energy-momentum tensor. If we choose $\beta = 0$ we go  ahead with the calculation and find the different components of $T_{\mu\nu}$. However, we do not know {\it a priori} whether the resulting energy-momentum tensor will satisfy appropriate physical conditions. Here comes a practical motivation to introduce the conformal factor. Allowing $\beta \neq 0$ we add a degree of freedom that can be used to satisfy some physical requirement as isotropy in the pressures or some equation of state.

Thus, the physical conditions imposed on $4D$ matter, like the energy conditions, are the ones that ultimately should determine $\beta$ and, therefore, the physical metric. 

The physical metric obtained in this way can then be used to evaluate possible effects from the extra dimension. In particular, we can proceed to evaluate the four-dimensional invariants ${\cal F}_{\mu\nu}{\cal F}^{\mu\nu}$, $\epsilon^{\mu\nu\lambda\rho}{\cal F}_{\mu\nu}{\cal F}_{\lambda\rho}$ as well as the conserved four-vectors $J^{\mu}_{m}$ and $J^{\mu}_{e}$. All these are well-defined $4D$ quantities\footnote{We note that $\epsilon^{\mu\nu\lambda\rho}{\cal F}_{\mu\nu}{\cal F}_{\lambda\rho}$ and $J^{\mu}_{m}$ depend on the induced metric only. While ${\cal F}_{\mu\nu}{\cal F}^{\mu\nu}$and $J^{\mu}_{e}$ also depend on the choice of $\beta$.} invariant under transformations in $5D$. They should help\footnote{``In relativity, invariants are diamonds. Do not throw away diamonds!" \cite{Taylor and Wheeler}} in the interpretation and formulation of concrete predictions. For this we need to consider specific exact solutions.

The interpretation of $J^{\mu}_{m}$ and $J^{\mu}_{e}$ as, respectively, the four-vectors of magnetic charge-current density and electrical charge-current density is not mandatory. However, they are conserved $4D$ quantities regardless of their particular interpretation. 

The signature of the extra dimension can, in principle, be negative ($\epsilon = -1$) or positive ($\epsilon = + 1$). Which means that it can be spacelike or timelike, respectively. In the first case $\hat{F}_{\mu\nu}$, could be interpreted as the (generalized) electromagnetic tensor. However, timelike extra dimensions are also discussed in the literature. In particular they can lead to the existence of multidimensional objects, called ``T holes" \cite{Bronnikov and Melnikov}.

Ordinary matter is mainly represented by $T_{\mu\nu}^{(I)}$. This tensor shares a common property with $T_{\mu\nu}^{(IV)}$, for the field of electromagnetic type. Namely, that both depend explicitly on the signature of the extra dimension. However, they have different behavior with respect to scale changes. Rescaling the scalar field $\Phi \rightarrow K\Phi$ induces a change $T_{\mu\nu}^{(I)}\rightarrow (1/K)^2T_{\mu\nu}^{(I)}$ and $T_{\mu\nu}^{(IV)}\rightarrow K^2T_{\mu\nu}^{(IV)}$. The scaling of the metric $\hat{g}_{\mu\nu}\rightarrow e^{2\beta}g_{\mu\nu}$ induces a change $T_{\mu\nu}^{(I)}\rightarrow e^{2\beta}T_{\mu\nu}^{(I)}$ and $T_{\mu\nu}^{(IV)}\rightarrow e^{-2\beta}T_{\mu\nu}^{(IV)}$. The first feature seems to indicate is that the effects from the extra dimension become more prominent in the presence of fields of the electromagnetic type rather 
than in ordinary matter. The second one, for positive $\beta$ means that at large distances ordinary matter dominates over electromagnetic-type fields. For negative $\beta$, the corresponding interpretation is that as we go to smaller regions, the fields prevail over matter. 

In summary, in this work we have provided the most general expression for the energy-momentum tensor compatible with the requirement of gauge invariance, instead of cylindricity. We have considered the most general expression for the physical metric as being conformal to the induced one. The choice of this factor should be limited by physical conditions imposed on the $4D$ energy-momentum tensor. We have found two four-vectors that separately satisfy the equation of continuity in $4D$. If the antisymmetric tensor $\hat{F}_{\mu\nu}$ can be identified with the electromagnetic one, then the theory predicts the existence of classical magnetic charge and current.

\end{document}